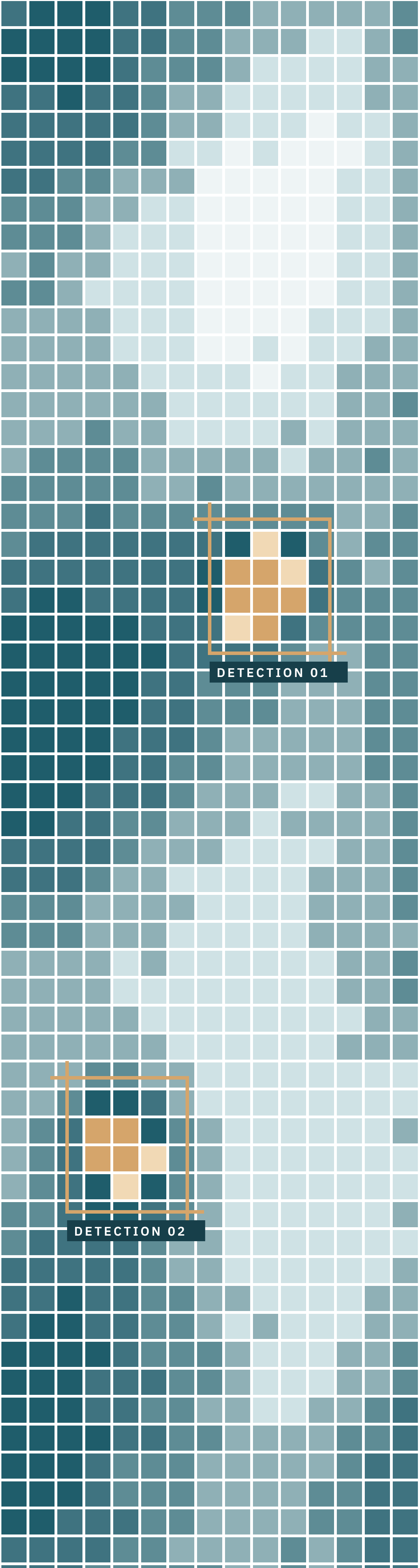

STEADFAST

RESEARCH REPORT

# Crisis Governance at the Joints

## Mapping the Ethical and Societal Territory of Crisis Management Technologies: The Case of Human-Swarm Teaming

AUTHOR:

**Dr. Anthony Longo**

Tilburg University

a.n.longo@tilburguniversity.edu

# Abstract

Crisis and disaster governance increasingly relies on technologies that sense, classify and act faster than humans can. Drone swarms coordinated with human operators, known as human-swarm teaming (HST), provide a productive case to review the field of crisis governance. They promise better situational awareness (SA) in disaster response, but they also raise questions about human control, responsibility, surveillance, triage, dual use and public legitimacy. This report maps the ethical and societal territory that HST opens up. It unpacks this territory with a "3i" framework: the Interface, where operators and swarms work together; the Infrastructure, where organisations distribute tasks, decisions and blame; and the Imaginary, where law, policy and shared societal assumptions define what a crisis is and which interventions it justifies. The central claim is that these concerns cannot be addressed one level at a time. Many of the most pressing problems, such as responsibility gaps and the erosion of meaningful human control, arise at the joints between levels, through four translations: habits harden, routines become stories, principles become procedures, and procedures become prompts. Drawing on a critical review of literature across philosophy of technology, political theory, moral psychology, philosophy of mind, AI ethics, and critical disaster and security studies, the report develops this framework, illustrates it with two worked examples, and derives design and governance implications and a set of questions for developers, emergency organisations and policymakers.

**About this report.** This report is a revised version of the first deliverable of Work Package 4 (Ethical, Legal and Societal Aspects) of the STEADFAST project. The original was written for the project consortium; this version addresses a wider audience. The views expressed are those of the author.

**Acknowledgements.** I thank Maureen Sie, Colette Cuijpers, Richard Heersmink, Lonneke Peperkamp, Marie Šafář Postma, and the other STEADFAST consortium and its Advisory Board for their feedback on earlier versions of this report. I used Claude (Anthropic / Opus 5.5) to design the figures 1, 3 and the cover image.

**Funding.** This publication is part of the project STEADFAST - Swarm Technology Enabling Advanced Drone-Facilitated Active Support Tactics for Military and First Responder Operations, with file number KICH1.VE04.22.004, of the research programme "Collaboration between humans and (semi-)autonomous systems" under the Dutch Knowledge and Innovation Covenant (KIC) grant https://doi.org/10.61686/SVBGA49631, which is (partly) financed by the Dutch Research Council (NWO).

**Suggested citation.** See SSRN guidelines.

# Table of Contents

# Executive Summary

Drone swarms promise to give emergency responders eyes where humans cannot go. This report asks what else they change. It treats human-swarm teaming (HST) as an entry point into the wider ethics of technology-mediated crisis governance. Its main messages are:

- **Problems live at the joints.** The hardest ethical problems of HST do not sit at one level. They arise where interface design, organisational relations and public imaginaries meet and translate into each other.
- **The interface is not a transparent window.** What operators see is already filtered, fused, scored and prioritised. Design choices about visualisation are morally sensitive choices about what operators can notice and care about.
- **A human in the loop is not enough.** Meaningful human control requires that people can understand, contest and answer for what the system does, across design, deployment and review.
- **Responsibility must be governed, not assumed.** Without explicit role-based accountability, blame tends to land on the operator nearest the system: the moral crumple zone.
- **Visibility is not the same as care.** If being visible to the system comes to mean being eligible for rescue, needs that are hard to measure drop out of disaster response.
- **Dual use is a design question.** Data, algorithms and categories migrate between civil and military uses. Purpose limitation has to be built into architectures and governance from the start.

## Who this report is for

| If you are... | Start with | Key concerns |
|---|---|---|
| A designer or engineer of swarms or crisis management technologies | Sections 3.1 and 4 | Technological mediation<br>Uncertainty display<br>Moral deskilling<br>Contestability |
| An emergency or defence organisation | Sections 3.2 and 6 | Responsibility gaps<br>Meaningful human control<br>Accountability<br>Ethics of triage |
| A policymaker or regulator | Sections 3.3 and 6 | Crisis imaginaries<br>Dual use<br>Public legitimacy |
| A researcher in ethics, STS or disaster studies | Sections 2 and 4 | The relational framework<br>The four translations |

SECTION 1

# 1. Introduction: Why Crisis Technologies Demand Ethical Scrutiny

**Drone swarms** are collectives of (often small) unmanned aerial vehicles (UAVs) whose coordination is decentralised: each unit follows relatively simple local rules, yet together they exhibit complex collective behaviour. When combined with human operators and command structures, these swarms form Human-Swarm Teaming (HST) systems. Compared to single UAVs, swarms promise greater scalability, robustness, energy efficiency and resilience, making them attractive for both humanitarian tasks (disaster relief, search-and-rescue, mass-casualty triage) and military applications (intelligence, surveillance, reconnaissance, and potentially strike coordination).

This **dual-use** character of HST immediately raises ethical, legal, and societal concerns. Recent conflicts have shown that AI-enabled targeting and large-scale sensor networks can dramatically compress the time between data collection, target identification and action, often with limited human oversight and under conditions of radical uncertainty. At the same time, the features that make HST ethically contentious also underpin its appeal in civil protection: limited situational awareness has long been recognised as a central bottleneck in disaster zones, and HST architectures are designed precisely to address this.

In both civil and military domains, then, HST systems promise to extend human perceptual and decision-making reach. Yet this extension comes with significant **ethical, legal and societal risks**. As situational awareness is increasingly algorithmically mediated and partially autonomous, **questions** emerge about how trust, control, and responsibility are to be understood and distributed: Who is accountable when decisions are based on opaque sensor-fusion algorithms or adaptive autonomy mechanisms? How should responsibility be allocated when human operators supervise large swarms through combined data visualisations rather than direct control of individual units? Under what conditions can “meaningful human control” (MHC) be said to exist in systems that are explicitly designed to act faster than unaided humans could? These questions do not concern only individual operators; they also touch on how organisations, legal frameworks, and broader societal imaginaries of what “crisis”, “security”, and “safety” *are* evolve around swarm-based infrastructures.

This report was written within STEADFAST, a Dutch NOW-funded research project that develops and tests human-swarm teaming for military and first-responder operations, with a focus on situational awareness, adaptive autonomy and human-swarm interfaces. The project includes a work package on ethical, legal and societal aspects, and an earlier version of this report was that work package's first deliverable. This public version keeps the analysis but addresses a wider audience: anyone who designs, procures, deploys or governs technologies for crisis response, and anyone who studies them. Human-swarm teaming serves as an entry point. Many of the issues it raises apply equally to other sensing and decision-support technologies in disaster and crisis governance.

> **This report** maps the ethical and societal territory of technology-mediated crisis governance, using human-swarm teaming as its case. It makes **three contributions**:
>
> (1) First, it offers a **relational "3i" framework** with three levels (interface, infrastructure and imaginary) and four translations between them.
>
> (2) Second, it **maps the literature** at each level and shows that the most pressing problems, such as responsibility gaps and the erosion of human control, arise at the joints between levels.
>
> (3) Third, it derives **design and governance implications** and a set of questions for developers, emergency organisations and policymakers.
>
> **Method and scope.** The report is a critical, narrative review of literature across philosophy of technology, political theory, moral psychology, philosophy of mind, AI ethics, and critical disaster and security studies. It is not a systematic review. Its focus is civil disaster and crisis response. Because swarm technologies are developed for civil and military use alike, the boundary between the two is treated as part of the territory, not outside it.

SECTION 2

# 2. Ethics at the Joints: A 3i-Framework

Human-swarm teaming is not just a new tool for disaster response and crisis management. It is part of a wider sociotechnical infrastructure for sensing, deciding and acting in crisis. A framework that looks only at individual operators (for example, whether a person 'stays in control') or only at abstract principles (for example, 'human dignity') is therefore insufficient. Responsible development and deployment require an approach that shows how local design choices, organisational arrangements and broader legal-political imaginaries of "crisis" and "security" continuously shape one another.

The framework distinguishes three levels and stresses the relations between them. At **the Interface** (micro level), HST appears as concrete interactions between human operators and swarms: interfaces, autonomy modes, cognitive demands and situated judgments. At **the Infrastructure** (meso level), it appears as an organisational ecosystem: how roles, protocols, training, logging and evaluation are configured around the technology. At **the Imaginary** (macro level), HST is embedded in legal regimes, policy frameworks, ethical doctrines and wider societal imaginaries of war, peace, crisis and public safety. The framework is not a simple stack of layers. It is a way to trace how ethical and legal responsibility is translated, redistributed and sometimes diluted as practices and technologies move between levels.

This leads to the central claim of the report: **ethical concerns about crisis technologies cannot be addressed one level at a time.** Many familiar problems arise at the joints between levels. Section 2.2 describes the four translations through which this happens, and Section 4 follows two problems across all three levels.

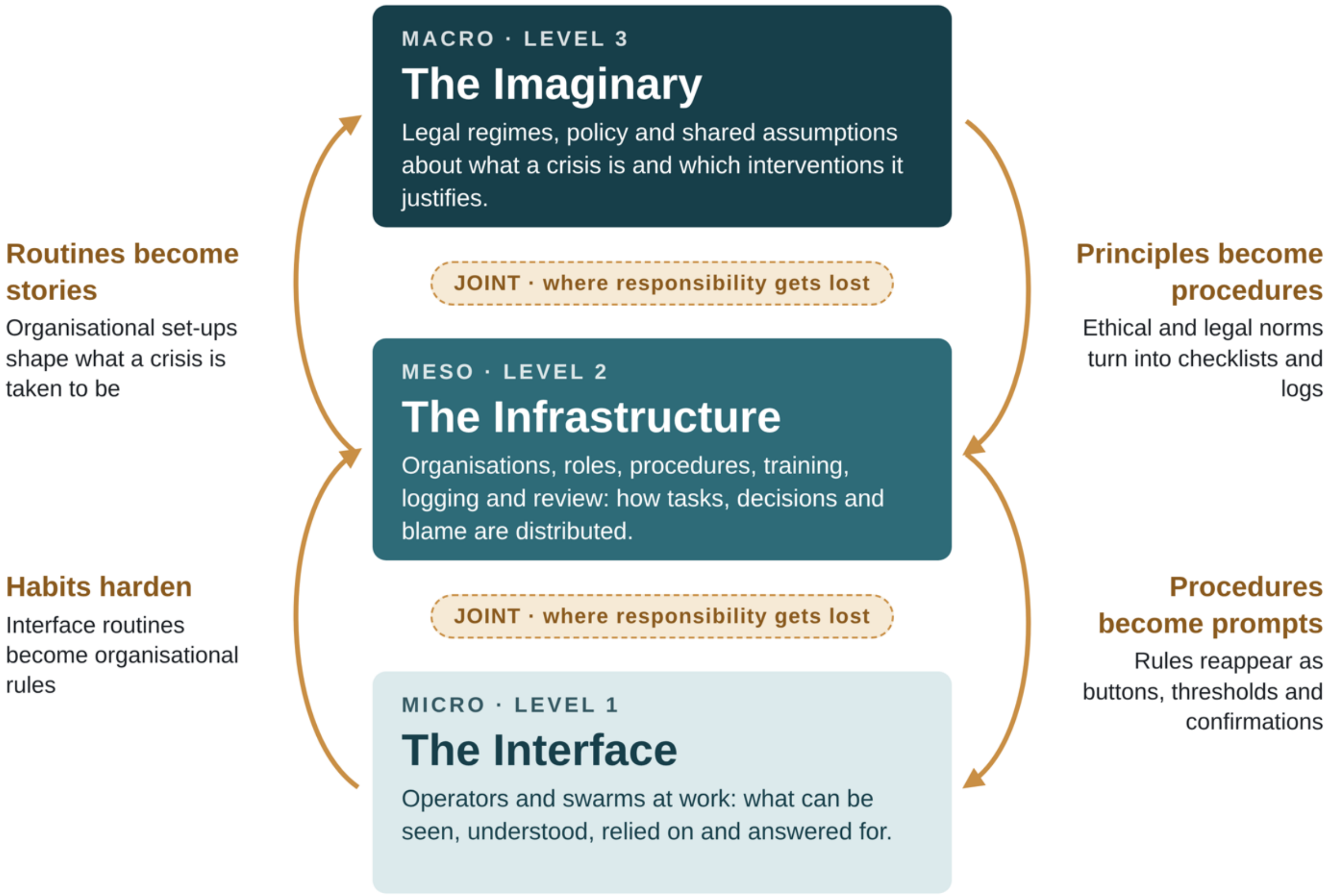

**Figure 1.** The three levels of the framework (the Interface, the Infrastructure and the Imaginary) and the four translations between them. Many ethical problems arise at the joints between levels.

## 2.1 The Three Levels

**The Interface (micro level):** The micro-level concerns the concrete situations in which humans and swarms jointly perform tasks: e.g. a mission commander supervising multiple drones in a search-and-rescue operation; an ISR operator monitoring a dense 3D visualisation of an urban space; a first responder interpreting real-time triage data during a mass-casualty event. In projects such as STEADFAST, this level is the focus of work on human-swarm interfaces, adaptive autonomy based on the operator's cognitive state, sensing and data fusion, where the aim is to improve situational awareness under high cognitive and temporal pressure.

Ethical questions at this level typically take the form: What can the operator see, understand, and legitimately rely on? How is attention steered by interface design? Under which conditions can operators intervene, override, or refuse system suggestions? How is responsibility experienced when decisions are mediated by complex sensor-fusion, predictive models, and adaptive autonomy mechanisms? Here, key concepts such as human control, trust in automation, cognitive load, moral agency, and “response-ability” are examined not in the abstract but in relation to concrete design choices and interaction patterns.

**The Infrastructure (meso level):** The meso-level shifts the focus from individual operators to the organisations and infrastructures within which HST is embedded. This includes the way armed forces, air commands, crisis-management centres, and first-responder organisations design roles, relations and hierarchies around HST; the standard operating procedures and rules of engagement that define who may authorise what; the training and certification practices that determine how operators are prepared; and the logging, audit, and review systems that document what happened and why. In projects such as STEADFAST, this level is addressed in roadmaps for implementation, in training and education, and in co-creation with air forces, defence partners and civil first responders.

Ethical and legal issues at this level concern the distribution and stabilisation of responsibility: how tasks, risks, and decision rights are allocated across roles; how after-action reviews and incident investigations assign praise, blame, or calls for redesign; how doctrines and procedures encode particular values (e.g. speed versus deliberation, resilience versus transparency). The meso-level also includes the “infrastructures of responsibility” that organisations build: who has access to which data, when escalation is required, what must be logged and justified, and how operators are appraised over time. These arrangements can either support or erode meaningful human judgment, contestation, and learning.

**The Imaginary (macro level):** At the macro-level, HST is situated within broader legal, ethical, and political frameworks. This includes international humanitarian law (IHL), human rights law, Dutch and EU regulations on AI, autonomous systems and data protection, and emerging NATO and national doctrines on responsible military AI. It also includes public concerns with surveillance, privacy, and the militarisation of AI, as well as more diffuse “crisis imaginaries”: i.e. shared assumptions about what crises are, how they unfold, and what forms of exceptional intervention they justify.

An ethical stance at this level requires engagement with questions such as: What underlying conception of ‘responsibility’ drives the hypothesis that HST can be designed and used responsibly across civilian and military use cases? How do legal categories such as “emergency”, “exceptional measures”, or “battlefield” relate to complex urban crises, grey-zone conflicts, or dual-use deployments? What narratives about risk, resilience, and preparedness are mobilised to legitimise new infrastructures such as drone swarms, and how do they affect democratic oversight and public trust? How can research projects play a productive role in shaping responsible narratives about HST in disaster response?

| Level | What it covers | Typical questions | Examples |
|---|---|---|---|
| The Interface (micro) | Concrete situations in which humans and swarms jointly perform tasks: interfaces, autonomy modes, cognitive demands, moral perception. | What does the operator actually see and understand?<br>When can they intervene, override, or refuse?<br>How is responsibility experienced? | Physiological (e.g. EEG-based) cognitive state estimation;<br>Swarm control dashboards and visualisations;<br>Adaptive autonomy shifting in real time;<br>Cognitive load monitoring |
| The Infrastructure (meso) | The organisations, doctrines and infrastructures within which HST is embedded: roles, training, standard operating procedures, logging systems, review practices. | How are tasks and decision rights allocated across roles?<br>How do after-action reviews assign responsibility?<br>What must be logged and justified? | Air force and fire-service command structures;<br>Automated mass-casualty triage protocols;<br>Audit-trail and replay systems |
| The Imaginary (macro) | Legal regimes, policy frameworks, ethical doctrines and societal imaginaries of disaster, crisis and public safety. | What narratives about risk justify new HST infrastructures?<br>What are the dual-use implications? | Dutch & EU regulatory frameworks for UAS and AI;<br>NATO AI governance;<br>IHL compliance in military;<br>Public trust in BVLOS operations;<br>Public communication |

## 2.2 How the Levels Connect: Four Translations

The levels are not stacked on top of one another. They are linked by four translations. These matter because many familiar ethical problems, such as responsibility gaps and the erosion of human control, arise at the joints between levels rather than within any single level. Each translation is given a plain name below, with the more technical term in the text.

### Habits harden (Interface → Infrastructure)

In what can be called the sedimentation of practices, local interaction patterns and design choices at the operator–swarm interface gradually solidify into organisational relations and routines. An interface that prioritises certain alerts, or an adaptive autonomy mode that reallocates tasks when workload is high, can become a ‘normative’ and ‘authoritative’ template for how work is *supposed* to be done. Roles are redefined around what the technology/system presupposes (operators as supervisors, commanders as validators), and responsibility shifts to role-based accountability.

> **Why it matters:** Interface design decisions made in development projects become embedded, over time, in standard operating procedures and training. Early choices about what information is surfaced and what is suppressed shape organisational responsibility practices for years.

### Routines become stories (Infrastructure → Imaginary)

Once stabilised, organisational configurations inform broader narratives about disaster, crisis, security and public safety. Legal and policy categories may then start to mirror these assumptions, with emphasis shifting from public justification after the fact to anticipatory risk governance. The long-term outcomes of HST research projects are therefore not limited to direct ‘products’; they may also contribute to the changing societal image of what a ‘disaster’ is and what kind of ‘treatments’ it legitimizes.

> **Why it matters:** Implementation roadmaps and pilot projects can shape national and European narratives about what responsible HST deployment looks like. How these narratives are framed determines whether exceptional measures become normal.

### Principles become procedures (Imaginary → Infrastructure)

Macro-level principles (e.g. 'meaningful human control', IHL and EU AI Act requirements) do not apply themselves easily. They are translated into concrete governance templates: audit trails, escalation ladders, compliance checklists, oversight bodies. These translations often encode a specific vision of ‘responsibility’ as something that can be secured *procedurally*, e.g. through logs, reviews and formal authorisations, rather than exercised substantively through open-ended judgment.

> **Why it matters:** Transparency, traceability and explainability requirements should be evaluated not only for technical compliance, but for whether they actually support human judgment or merely demonstrate control on paper.

### Procedures become prompts (Infrastructure → Interface)

Organisational and governance infrastructures materialise in the everyday experience of operators. Oversight requirements become dashboard prompts; legal accountability becomes thresholds that must be confirmed; after-action review practices shape how operators anticipate later appraisal. Operators may retain formal responsibility but lose the practical conditions for meaningful interruption, dissent or alternative interpretation.

> **Why it matters:** Work on cognitive load and adaptive autonomy should be evaluated not only with performance metrics, but by whether it sustains or erodes the operator's capacity for moral judgment.

SECTION 3

# 3. Mapping the Territory: Ethical and Societal Aspects of HST

The following three subsections map the ethical and societal literature at each level of the framework. The review is critical and narrative rather than systematic: it selects literature for its relevance to the questions HST raises, across philosophy of technology, AI ethics, human factors, and critical disaster and security studies. Each subsection closes with preliminary design and governance implications. These are working hypotheses, not final recommendations.

THE INTERFACE Level 1 of 3 · operator–swarm relations (micro)

## 3.1 The Interface: The Ethics of Operator–Swarm Relations

At the Interface (the micro level), the ethical questions surrounding Human–Swarm Teaming (HST) revolve around the human operator's **cognitive, perceptual, and moral relationship** with the swarm. Human–Swarm interfaces create complex perceptual environments in which video feeds, radar data, lidar scans, predictive models, and sensor fusion outputs are continuously integrated. **The operator's task is not merely to decide, but to interpret**: to synthesize heterogeneous signals into a meaningful situational picture while under conditions of time pressure, uncertainty, and moral weight.

As Schurr et al. (2005) demonstrated in early multi-agent systems research, limited situational awareness significantly reduces the effectiveness of human coordination. In HST, that challenge is amplified: humans must operate within dynamically shifting information ecologies in which both human and machine cognition co-evolve. The core ethical issue thus shifts from whether humans "stay in control" to whether they can *understand and respond meaningfully* to what is taking place.

### Technological Mediation

HST operators do not encounter the environment directly. They work within *mediated environments* built from layered thermal indicators, aerial maps, algorithmic predictions, confidence scores and risk colour codes. The moral significance of this can be articulated via **technological mediation**: technologies co-constitute moral perception by amplifying some affordances and attenuating others (Ihde 1990; Verbeek 2011). In HST, *amplification* might occur when thermal signatures of survivors are enhanced and visually prioritised; *reduction* occurs when areas of high uncertainty or contextual ambiguity are de-emphasised or smoothed into aggregate scores.

Together, these layers form an **evidential ecology** (Gillett and Heersmink 2019): a mediated field that defines not only what can be known and what counts as evidence, but also what can be cared about and acted upon. The ethics of HST begins long before any discrete decision point. It begins with how technologies make the world 'knowable'.

| Key concept | **Evidential ecology**: the mediated field produced by an HST interface that determines what is visible, what counts as evidence, and what courses of action are implied. Design choices about visualisation, alerting, and classification are simultaneously ethical choices about what operators can care about. |
|---|---|

### Interpretive Labour and Moral Responsibility

Within this evidential ecology, **interpretive gaps** inevitably arise. The meaning of a coloured icon, a temperature gradient, or a classification label is not self-evident; each encodes prior design decisions about which features matter and which can be treated as noise. Choices about map scale, iconography, thresholds for alerts, or the visual salience of certain categories (e.g. “threat”, “safe zone”, “civilian”) quietly steer attention, frame priorities, and imply default courses of action (Endsley 2015). By highlighting some features of a situation—movement patterns, heat signatures, anomaly scores—while downplaying others—emotional distress, ambiguity, or contextual uncertainty—the interface shapes both what the operator sees and what they can practically take into account when judging a situation.

Operating within a swarm therefore requires a distinctive form of **interpretive labour** that goes beyond traditional models of “decision-making”. Operators must maintain situational awareness, integrate heterogeneous signals, track the swarm's evolving state, and exercise moral discernment in the face of algorithmically produced suggestions and ambiguities (Hocraffer and Nam 2017). Counterintuitively, it is long known that automation can *increase* rather than reduce cognitive demands: even if some tasks are offloaded to algorithms, the need to monitor autonomous behaviour, interpret opaque outputs and detect failures can generate high cognitive load (Parasuraman and Riley 1997). Studies of multi-agent visualisation systems such as DEFACTO suggest that 3D and multimodal displays can improve shared awareness and coordination (Schurr et al. 2005), but they also demonstrate that this coherence is provisional and heavily dependent on design choices; when AI components “fill in” perceived gaps, interpretive authority can quietly shift from humans to systems.

In this sense, **moral responsibility** is not simply a pre-given property of individual operators; it is shaped by the roles, affordances, and virtues that technologies enable/disable (Vallor 2016). The operator’s capacity to perceive others as vulnerable, to recognise morally relevant features, or to question normative assumptions is mediated through interface architectures, alert logics, and background models. At the same time, “moral patients” (i.e. beings that receive/deserve moral consideration) are constituted through how they appear—or fail to appear—within the system: as heat signatures, moving blobs, classified anomalies, or indistinct background noise. Ethical analysis at the micro-level thus requires close attention to how design decisions structure visibility, invisibility, and the kinds of moral questions that can arise within HST operations.

## From Control to Response-Ability

Dominant ethical frameworks tend to **equate responsibility with “control”,** presupposing a highly demanding moral psychology: a human agent who can always intervene knowingly, deliberately, and with full situational awareness (Santoni De Sio and Van Den Hoven 2018; Verdiesen et al. 2020). This **idealised picture** is both empirically unrealistic and normatively counterproductive in complex sociotechnical systems. Drawing on Sie’s “traffic-participation view” of moral agency (TPV), we can instead understand responsibility in terms of minimal conditions for competent participation in a norm-governed practice (Sie 2009, 2014, 2018). On this view, responsible agency does not require continuous conscious control or explicit deliberation; rather, it requires that agents are generally able to recognise relevant reasons, respond appropriately within their roles, and be held answerable through meaningful practices of appraisal and feedback over time.

Applied to HST, this perspective shifts the question from “Does the operator stay in control?” to “Do organisational and technical infrastructures sustain **minimal responsibility competence** and meaningful appraisal practices?”. Organisational and governance structures materialise at the micro-level in the roles and interfaces that operators encounter: oversight requirements become dashboard views and confirmation prompts; legal accountability becomes specific thresholds that must be acknowledged or logged; escalation procedures become buttons or workflows embedded into the interface. These configurations pre-structure how responsibility is experienced: more as an obligation to follow, supervise, and confirm processes designed elsewhere than as a space for substantive judgment.

**Responsibility is enacted *longitudinally*** (i.e. over time) through “tokens of appraisal” (e.g. praise, blame, correction, after-action reviews, and incident investigations) rather than at isolated decision points (Sie 2015, 2018). In HST contexts, many of these appraisal mechanisms are tightly coupled to technical infrastructures: logging systems, replay timelines, performance dashboards, and automated notifications. Such mechanisms do normative work: they shape future attention, calibrate trust in automation, and influence operators’ willingness to intervene (Kudina 2021). They also offer differentiated responses to failure. Rather than oscillating between scapegoating individuals and resigning to “the system did it”, it is useful to distinguish three kinds of cases: competence failures or overload (“mad”), culpable norm violations (“bad”), and reasonable disagreements about the norms themselves ("disagreeing") (Sie 2000). Designing HST infrastructures to support this differentiation can help avoid both unjust blame and systemic moral abdication.

This implies that operators need not necessarily possess omniscient oversight to remain responsible, but they must inhabit systems that sustain their capacity to respond, what can be called **response-ability**. Response-ability shifts ethical focus from preserving command or control to cultivating the conditions under which humans can remain perceptually and morally attuned to others (human, technological, environmental aspects of the situation). Designing for response-ability thus means shaping interpretive environments that sustain judgment rather than replace it. It involves interface features that expose counter-evidence instead of concealing it, transparency about the operator's current role (for example, "you are now acting as gatekeeper"), and audit mechanisms that record reasons available at the time, not only outcomes. Training and feedback systems must also move beyond punishment toward 'tokens of appraisal': practices that reinforce and reward sensitivity to reasons and expand moral awareness under pressure.

| Key concept | **Response-ability**: the capacity of operators to remain perceptually and morally attuned to their situation within an HST system. Designing for response-ability means shaping interfaces and training environments that sustain perception and judgment rather than replace it. |
|---|---|

## Adaptive Autonomy and Cognitive Monitoring

Recent work in human-AI teaming adds another layer of ethical complexity through the use of **adaptive autonomy**. Here, systems adjust the level of swarm autonomy dynamically in response to estimates of the operator's cognitive state, derived from psychophysiological measures such as EEG, heart rate variability, respiration, and eye movement (Dehais et al. 2020; Charles and Nixon 2019). Studies in aviation, human–robot interaction, and neuroergonomics suggest that such monitoring can help modulate workload and prevent overload or fatigue by shifting tasks between human and machine in real time (McNeese et al. 2018; Hauptman et al. 2024). In line with this, research projects such as STEADFAST explore how cognitive state estimation might be used to support flexible, adaptive autonomy for HST operators.

While promising from a performance and efficiency perspective, such continuous monitoring raises four ethical concerns:

1) **Surveillance vs. support:** Continuous monitoring of cognitive and affective states blurs the boundary between wellbeing support and workplace surveillance. We can reasonably anticipate that well-intended cognitive monitoring measures may undergo "function creep": data collected to detect fatigue, prevent overload, or improve safety may later be used for productivity ranking, promotion, disciplinary matters, or other managerial decisions. Workplace hierarchies make it difficult to obtain truly voluntary informed consent. An ethical risk is that cognitive monitoring will transform supportive intervention into an oppressive surveillance infrastructure (Muhl and Andorno 2023; Chowdhary et al. 2023; Awumey et al. 2024).
2) **Asymmetric transparency:** The human operator becomes increasingly legible to the system through measurement, while the system's own opaque adaptive logics may remain difficult for the operator to inspect. Operators may not understand why autonomy levels shift, which metrics trigger them, or how their bodily responses are being classified. Such a persistent asymmetry may cause operators to become over-reliant on the system, lose situational awareness, or be unable to intervene meaningfully (Chen et al. 2018; Zerilli et al. 2019; Hussein et al. 2022).
3) **Moral deskilling:** Adaptive autonomy may produce moral deskilling when it removes operators from the difficult, context-sensitive cases through which practical judgment is developed and maintained. Practical wisdom depends on repeated engagement with particular situations; delegating such decisions can weaken human competence altogether and encourage responsibility offloading. In HST, passive monitoring and restricted intervention have been associated with reduced SA, perceived control, and readiness to intervene. This risk is especially serious where operators remain morally and legally responsible for decisions they are increasingly trained not to make. Adaptive autonomy should therefore preserve and train meaningful human involvement rather than treating human judgment as emergency backup (Vallor 2015; Eisikovits and Feldman 2022; Tsai and Ku 2025).

4) **Neurodivergence and disability:** Cognitive-state estimators often treat physiological signals as proxies for a supposedly general baseline of attention, engagement, stress, or competence. Yet neurodiversity scholarship cautions against interpreting atypical cognitive, sensory, attentional, or social-perceptual patterns as deficits rather than as forms of human variation. Models trained on neurotypical operators may misclassify atypical sensory and cognitive patterns, triggering unnecessary autonomy reduction, intensified supervision, or adverse fitness judgments (Newman-Griffis et al. 2023; UNESCO IBC 2023; Baeyaert 2025; Bariffi 2026)

## Moral Injury

A **common intuition** holds that added distance and reduced control should protect operators from psychological harm: if BVLOS operation removes physical danger, and adaptive autonomy removes operators from the immediate causal chain of harmful outcomes, moral-psychological risk should fall accordingly. This assumption underlies prevailing extensions of Just War Theory that ground a duty to deploy remote and autonomous systems partly in their expected capacity to *shield* operators from guilt, trauma and moral failure.

However, empirical research on remote drone operators challenges this directly. ***Physical* removal does not reliably produce *moral-psychological* distance**. On the contrary, multimodal sensor dashboards can generate heightened rather than diminished intimacy with the people (victims or offenders) being observed or tracked. HST interfaces may produce such intimate connections where operators see human-biological data up close while, at the same time, being removed from direct modes of care and thereby stripped of the “moral permission” that combatants/operators usually experience when facing direct danger (Enemark 2024). Thus, distance and mediation can compound moral weight rather than dissolve it.

This yields distinct risks for operators’ well-being on three levels (Molendijk et al. 2026): (1) while being immersed in a constant flow of information can inflate an operator’s **illusion of control**, disruptions (e.g. jammed sensors, missing/contradictory ground-level information) can leave operators particularly powerless and unconfident, often at the moment they most need to feel capable of acting; (2) operators can also experience **moral failure** and “tragic remorse” even where no procedural violation occurred and no responsibility is formally assigned to them; and (3) where operators come to doubt the fairness of the criteria embedded in the systems they operate, this can escalate into **general distrust** of the organisation that asked them to rely on those systems.

| **Design note** | Systems that use cognitive monitoring and adaptive autonomy should be evaluated not only for task performance and moral deskilling, but also for operators' moral-psychological well-being under sustained exposure. Training and support should account for this risk rather than assume that remote, swarm-mediated operation is protective. |
|---|---|

### Design Implications at the Interface

| Implication | What this means in practice | Who should act |
|---|---|---|
| **1. Design interfaces to make uncertainty visible** | Visualisations should surface conflicting data, ambiguous zones and the limits of model confidence—not smooth them into aggregate scores or single colour codes. Operators should be invited to exercise interpretive judgment. | Interface designers; developers of sensor fusion |
| **2. Ensure transparency and contestability in adaptive autonomy** | Operators must be able to understand when and why autonomy levels shift, and must have accessible means to contest or override these shifts. Cognitive monitoring must be subject to clear norms of consent, data use and privacy protection. | Developers of adaptive autonomy; operator organisations; data protection officers |
| **3. Support longitudinal responsibility practices** | Logging and replay systems should be designed not only for forensic blame-allocation, but for learning and norm refinement. They should facilitate after-action reviews that distinguish competence breakdowns, culpable violations and reasonable mistakes. | Developers of logging systems; trainers; operational commanders |
| **4. Scaffold cognition rather than deskill** | HST systems should be evaluated for the epistemic environment they create: do they *scaffold* understanding and reason-giving, or do they *reduce* human contributions to mere confirmation? Do they allow/disable operators to train and cultivate their capacity of judgment? | System designers; evaluators; procurers |

**THE INFRASTRUCTURE** Level 2 of 3 · organisational ecosystems (meso)

## 3.2 The Infrastructure: The Ethics of HST Ecosystems

At the meso level, ethical analysis shifts from the individual operator to the **collective ecosystem** of actors, interfaces and organisational arrangements that constitute HST. The central question is not only how individual humans act within the swarm, but how the human-swarm *team* acts as a sociotechnical system, distributing agency, decision-making and moral responsibility across organisational roles, procedures, training regimes and infrastructural configurations.

### Responsibility Gaps

As AI systems acquire increasing autonomy, speed, and opacity (particularly in high-stakes environments such as disaster triage or military engagement) concerns arise that traditional structures of accountability may erode. Traditional conceptions of moral responsibility rest on a unified and distinct picture of agency: intention > conscious decision > act > consequence. This view allows to trace back responsibility to a single deliberating subject. Autonomous AI systems disrupt this chain by distributing agency across a **complex field of decision-making**: concept, system design, target function, data, training, model, validation and testing, use, optimization, up to legal frameworks.

Because no single actor in this chain intends or decides in the way the traditional model presupposes, each party can plausibly deflect responsibility onto another: designers point to having only built an early framework, data scientists say they merely selected optimal metrics, quality assurance notes the system met all requirements, implementers claim they followed protocol, management says it relied on expert advice, lawmakers point to certification, and operators note the system acted autonomously—while the system itself cannot answer at all. This results in a **"responsibility gap"** (Matthias 2004): moral harm clearly occurs, but no single link in the chain can be rightfully identified as bearing blame for it. de Sio and Mecacci refine Matthias' analysis by distinguishing **four distinct responsibility gaps** (Santoni de Sio and Mecacci 2021):

| Gap | Responsibility type | Problem |
|---|---|---|
| *Culpability gap* | Passive (backward-looking) | No human can be legitimately blamed for harm |
| *Moral accountability gap* | | Agent(s) can't explain the "why" of a decision to others |
| *Public accountability gap* | | Citizens get no explanation from public agents/institutions |
| *Active responsibility gap* | Active (forward-looking) | No human is sufficiently aware, capable, or motivated to *prevent* harm in the first place |

A nominal human supervisor does not close these gaps. Effective oversight requires causal-effective power, epistemic access, self-control, and fitting intentions (Sterz et al. 2024). Otherwise, the operator becomes a **moral crumple zone**: the person nearest the system absorbs blame for failures produced by a wider socio-technical arrangement (Elish 2019). Human-in-the-loop requirements may even encourage "rubber-stamping" and shift responsibility from designers and institutions to frontline personnel (Green 2022). Under these conditions, assigning responsibility becomes both normatively urgent and practically difficult.

## Meaningful Human Control

The concept of Meaningful Human Control (MHC) has emerged as the principal ethical-legal strategy for preserving responsibility in these conditions, though sometimes Appropriate Human Judgment (AHJ) is preferred. This resonates with the EU AI Act, which classifies AI systems used for emergency dispatch and emergency patient triage, and safety components of critical infrastructure, as high-risk, and requires human oversight for such systems.

MHC does not simply require that a human is 'in the loop.' Rather, it reconceptualises responsibility as a property of sociotechnical systems designed to remain responsive to human-moral reasons. Systems must be designed so that human agents can meaningfully shape outcomes in accordance with morally relevant facts and reasons—inspired by Fischer and Ravizza's (1998) notion of 'guidance control' (Santoni de Sio & Mecacci 2021). Two conditions arise for MHC:

**(1)** **Tracking** requires that system behaviour responds to relevant human moral reasons and relevant facts in the world. A swarm must not merely execute a technically defined objective; it must also remain responsive to the reasons and facts that make the mission legitimate. This includes: (1.1) *world-tracking*: representing relevant environmental facts accurately enough for the task; and (1.2) *reason-tracking*: responding to the values, purposes, and changing instructions of responsible humans.

**(2)** **Tracing** requires that system actions be linked to a proper moral and technical understanding on the part of relevant humans in the design or use chain. At least one responsible human or institution must understand, or be in a position to understand, the system's capabilities, possible effects, and role in producing them.

| Guidance Control (GC) | | Meaningful Human Control (MHC) |
|---|---|---|
| **1) Reason-responsiveness** | = DM can recognize and follow up on good reasons to act differently<br>= sensitivity towards moral reasons and relevant facts | **1) Tracking:** DM co-varies *(tracks with)* relevant moral reasons and facts<br>=> decide relevant moral reasons<br>=> include relevant sensors (fact-tracking) |
| **2) Ownership** | = you take ownership of the DM + you know and accept that others can hold you accountable<br>= you understand the DM and you support it | **2) Tracing:** every outcome can be traced to at least 1 person with understanding and ownership of DM<br>=> design for action logging<br>=> clear role and competence distribution<br>=> technical literacy |

**Figure 2.** From the conditions of guidance control (GC) to the tracking and tracing conditions of Meaningful Human Control (MHC), based on Santoni de Sio & Mecacci (2021). *DM* = decision-making mechanism.

The sometimes preferred term of **'Appropriate Human Judgment' (AHJ)** can be understood as the role-sensitive application of human judgment where contextual understanding, normative choice, exception handling, or responsibility-bearing cannot safely be delegated. It does not imply that human judgment is always superior to automated performance. It implies that the **division of labor must reserve the right kinds of judgment to humans**. For instance, in disaster response, a swarm may autonomously search, map, or detect hazards, while human operators retain the *capacity and authority to judge* over rescue priorities, acceptable risks to responders, privacy-sensitive uses, and exceptional cases (Verhagen et al. 2024). In military ISR, humans should retain judgment over collection purposes, target-relevant categories, uncertainty, dissemination, and the downstream significance of machine-generated classifications (Ekelhof 2019; Verdiesen et al. 2020). This requires the team to distinguish moral (human) tasks from technical (machine-delegated) tasks.

Some scholars have argued that MHC is incomplete without **contestability** mechanisms (Alfrink et al. 2022; Lyons et al. 2021). A contestable system allows operators to challenge unreliable inputs, distribution shifts or changing autonomy levels, missing features, bias, or spurious correlations between 'defeaters' (Veluwenkamp and Buijsman 2025). It allows operators, affected communities and public bodies to challenge mission purposes, privacy impacts, exclusion risks, systemic errors and institutional governance. It also explicitly turns MHC into a **longitudinal learning process**: a challenge may reveal that the system's contexts, goals or saliency structure is wrong.

In sum, MHC does not require a human to direct every drone movement. It requires human and institutional control over mission purpose, operating boundaries, morally significant classifications, escalation, intervention where feasible, review, and accountability. This is compatible with bounded autonomy for navigation, mapping, search, and data fusion.

| Key concept | **Meaningful Human Control (MHC) / Appropriate Human Judgment:** not simply 'human in the loop,' but a property of sociotechnical systems designed to remain responsive to human moral reasons across design, training, deployment and review phases. For drone swarms, this is necessarily distributed and multi-layered, not located at a single point of operator intervention. |
|---|---|

## Distributed Responsibility in Practice

Recognising the limits of individual-centred approaches, recent scholarship proposes models of **distributed responsibility** tailored to complex AI systems. Rather than attempting to restore traditional models of *individual* accountability, these frameworks accept that responsibility in sociotechnical ecosystems must be shared, traceable and actionable across *multiple* actors and phases (Strasser 2021).

Ekelhof's (2019) analysis of targeting practices is instructive here: even in non-autonomous military AI systems, human agency is typically fragmented across the Find-Fix-Track-Target-Engage-Assess (F2T2EA) cycle. No single actor exercises full control at any point. For HST, this means that distributed responsibility is not a novel challenge posed by autonomy; rather, it is an extension of already-existing military practice. The question is how to *design systems and governance structures that* ***make this distribution explicit*** *(i.e. manageable and overseeable),* ***traceable*** *and ethically* ***coherent***, rather than leaving it implicit, arbitrary and unmanaged.

Applied to HST, distributed responsibility might materialise as **role-sensitive accountability matrices**: responsibility should be commensurate with actual ability and authority (Cavalcante Siebert et al. 2023). This requires explicit mappings linking swarm actions back to the humans and processes influencing them. In simplified form, and by manner of example, this *may* look like this:

| Role | Primary responsibility | Accountability mechanisms |
|---|---|---|
| **Interface engineers** | Design HST interfaces that support faithful communication of uncertainty levels, accessible data visualizations, and appropriate human judgment where needed (epistemic responsibility) | Design reviews; testing and validation; documentation; technical audits |
| **Human operator** | Monitor swarm behaviour, interpret system outputs, decide when to intervene or override, respect protocols and ethical codes (operational responsibility) | Mission logs; intervention records; after-action reviews and feedback |
| **Mission commander** | Define mission objectives in line with ethical-legal principles, operational constraints and acceptable risk thresholds, and authorise deployment (strategic responsibility) | Command authorisation procedures; operational oversight; mission debriefs |
| **Organisation / management** | Establish governance procedures, training requirements, accountability & role distribution, continuous evaluation (institutional responsibility) | Standard operating procedures; governance boards; compliance reviews; organisational audits |

## Transparency, Traceability and Explainability (TTE)

Distributed responsibility is impossible if no actor can understand how the swarm behaved or how an autonomous decision was produced. Legibility does not require complete access to every internal computation, but it does require sufficient, role-appropriate access to the information needed to act, explain, challenge and review.

The literature proposes three key design properties to enable MHC and distributed responsibility in practice:

**(1)** **Transparency** refers to the *accessibility* of system operations to meaningful human understanding: i.e. not low-level code inspection, but a workable grasp of how raw or synthetic data sources, proxies, algorithms and decision heuristics interact. They should nevertheless be treated as one layer of a broader transparency architecture: by themselves, they may not reveal distributed coordination, emergent behaviour, or the full provenance of a decision. Linking explanations to records of data lineage, model versions, transformations, and operational events is therefore important for accountability (Chen et al. 2018; Hepworth et al. 2021).

**(2)** **Traceability** focuses on the *post-hoc* capacity to reconstruct and audit behaviour, ensuring tamper-proof documentation of the swarm's decision trajectory. In high-stakes HST scenarios, whether combat reconnaissance or disaster triage, such mechanisms are indispensable for incident investigation and for distinguishing software failures from procedural breakdowns. Robotics research has proposed accountability architectures and ethical black boxes that use secure or tamper-evident logging to support behavioural reconstruction, explanation, and incident investigation (Fernández-Becerra et al. 2024; Winfield et al. 2025).

**(3)** **Explainability** targets contextually appropriate *reasons* for autonomous actions, empowering human teammates to anticipate and intervene effectively. Unlike raw transparency (which can overwhelm), explainability demands selective, role-sensitive, just-in-time communication: a dashboard might surface a concise rationale during operation (e.g. 'prioritised due to 85% thermal match and motion anomaly'), escalating to fuller decompositions post-mission (Özcan et al. 2025). Explanation without a reviewer or remedy creates a 'transparency fallacy' (Edwards and Veale 2018; Henin and Le Métayer 2021): dashboards that surface confidence scores and feature rationales may satisfy a formal explainability requirement while not giving operators the genuine capacity to verify, contest, or correct the underlying classification.

| **Design note** | Sensor fusion and data aggregation produce the outputs that determine whether transparency, traceability and explainability are achievable at the operator interface. Designing for explainability from the outset is far more tractable than retrofitting it. Developers and ethicists should agree early on what 'interpretable output' means for swarm recommendations. |
|---|---|

## Ethics of Triage

Triage is the practice of deciding who should receive attention, treatment or rescue first when needs exceed the available time, personnel or resources. It is therefore never merely a technical act of sorting people: it is a **morally consequential process** that distributes chances of survival, relief and recognition under conditions of uncertainty. Every triage system embeds assumptions about which signs indicate urgency, whose condition is most legible and which risks are acceptable. The Aerial Remote Triage System, for example, prioritises major bleeding, walking, consciousness and signs of life, while using one-way audio instructions and a "violet" category where death cannot ethically be classified remotely (Álvarez-García et al. 2021). Its protocol therefore embeds distributive judgements before any human supervisor acts.

The most urgent concern is that technical uncertainty can become **unequal access to care**. Existing studies and experiments remain prototype-heavy, often using healthy volunteers, simulated injuries and small samples. Drone-based heart-rate estimation produced an overall error of 14.3 beats per minute and failed in one-third of recordings, with performance affected by motion, exposure, occlusion, blood, dirt and skin tone (Mösch et al. 2023). Consciousness-detection systems similarly struggle with slight movement, clothing motion, thermal segmentation and changing environmental conditions (Pokee et al. 2021). Evaluation must therefore measure missed detections, unusable readings and priority-category reversals, and not only aggregate accuracy.

**Interface and protocol designs** are equally of ethical concern. Message prioritisation and visual salience determine which cases the operator/supervisor sees first (Diehl and Adams 2021). Human-AI teaming experiments in triage suggests that advice and task delegation preserve engagement and ethical compliance better than autonomous triage (Van den Bosch et al. 2025). The most advanced systems remain dependent on healthy volunteers, scripted behaviors, synthetic data, or small pilot datasets. The CERN ATTRACT Ethical Checklist appropriately treats trustworthiness as a socio-technical property requiring data provenance, lifecycle auditing, bias scrutiny, scenario fuzzing, and human oversight (Lee et al. 2026). Its **insistence on ethical hard stops** is especially relevant to mission drift: a system developed for casualty triage should not quietly become a surveillance or targeting system.

In addition, it is argued that **contestability** mechanisms must extend beyond the operator towards the **affected communities** (i.e. 'the triaged population'). Affected people need access to information about data sources, model performance and the division of diagnostic labour (Ploug and Holm 2021). Marginalised groups may also require intermediaries and collective support to contest decisions in order to promote inclusive citizen participation (Karusala et al. 2024). The central ethical principle is therefore: triage systems must be designed to make uncertainty, priority rules and avenues for human response visible to both supervisors and those being classified.

## Design Implications in the Infrastructure

| Implication | What this means in practice | Who should act |
|---|---|---|
| **1. Role-sensitive accountability mapping** | Develop explicit matrices linking swarm actions to responsible actors across the project lifecycle. Distinguish epistemic duties (engineers on interface design), practical duties (operators on intervention thresholds) and moral-political duties (commanders on mission parameters). Revisit these iteratively as HST evolves. | Emergency and defence organisations; system integrators |
| **2. Extend MHC beyond the operator interface** | MHC mechanisms should cover the full targeting/response cycle, embedding *upstream* validation (target vetting, collateral estimation) and *downstream* appraisal (after-action reviews) into standard operating procedures—not just the point of operator–swarm interaction. | Commanders; writers of doctrine and procedures; regulators |
| **3. Institutionalise TTE from the start** | Transparency, traceability and explainability should be built into data architectures from the outset—not retrofitted. Audit logs must support both real-time contestability and post-hoc reconstruction. Interface designs should balance informational completeness with cognitive load. | System architects; data engineers; auditors |
| **4. Design for organisational learning, not blame allocation** | Governance structures should prioritise learning-oriented appraisal over punishment, distinguishing competence failures, culpable violations and normative disagreements. After-action review practices should be embedded in training from the beginning. | Training institutes; incident investigators; management |

**THE IMAGINARY** Level 3 of 3 · legal-political and societal imaginaries (macro)

# 3.3 The Imaginary: The Ethics of Crisis Imaginaries

At the Imaginary (the macro level), ethical analysis shifts from the organisation of HST to the broader legal, political, and societal contexts in which HST is developed and deployed. The central question here is no longer how individual operators interact with swarms or how responsibility is distributed across an organisation, but **what *kind* of disaster risk management (DRM) HST participates in creating**. This broader perspective is important because HST does not just enter a neutral environment called "disaster". Disaster response itself is already structured by assumptions about preparedness, vulnerability, risk, resilience, visibility and legitimate intervention.

Technologies also *shape* how these terms are understood: what becomes visible, what counts as relevant risk, which interventions appear as 'necessary' and which forms of action become normal or legitimate. The macro-level thus raises questions about **'crisis-governance imaginaries'**, more specifically about how HST may transform (1) the meaning of 'crisis'; (2) the boundaries of intervention; (3) the values driving responsible innovation; and (4) the possibilities for democratic contestation and legitimization.

The concerns unpacked below should not be understood as predictions that HST will necessarily produce undesirable effects. Rather, they identify possible trajectories that should be made visible and deliberately governed during its development and implementation stages.

## Changing the Meaning of Crisis

One of the most important macro-level effects of HST concerns **what becomes visible and actionable as a 'crisis'**. A drone swarm does not merely move through a disaster environment; rather, through its sensors, models, communication architecture and task allocation systems, it helps organize what can be seen, measured, classified and acted upon. The use of aerial technologies across domains of conservation, humanitarianism, policing, and warfare—taken together—produce new forms of spatial visibility and intervention (Fish and Richardson 2022).

Critical voices distinguish between two modes of visibility: (i) **operational visibility** concerns whether a system can locate a person, hazard, vehicle or damaged structure; (ii) **political visibility** concerns *whose* situations become legible as a public problem, whose suffering attracts resources, and whose presence is interpreted as requiring intervention (Schnepf 2020). These two forms of visibility do not necessarily coincide: e.g. a drone may map flood damage with satisfactory accuracy while leaving questions about unequal protection, neglected housing or reduced public assistance outside the frame. The literature generally cautions that **faster technological detection does not automatically produce more inclusive or just disaster governance**. If increasingly large parts of a disaster environment can be sensed and automatically triaged, 'being visible to the system' may gradually become associated with 'being eligible for rescue'. Consequently, needs that are difficult to measure may become marginalized in DRM. The ability to see from above/aside will therefore not automatically lead to comprehensive 'situational awareness'.

This issue is particularly relevant to the ***humanitarian* framing of drones**. It is widely accepted that search-and-rescue (S&R) missions, damage assessment and last-mile assistance can provide genuine benefits, especially where human responders face dangerous or inaccessible environments. A scoping review of humanitarian drones identifies potential gains in harm reduction and welfare alongside concerns relating to justice, privacy, community relations, regulation and institutional identity (Wang et al. 2021). At the same time, humanitarian robotics can also alter relationships between responders and affected people (for the worse) by reducing contextual knowledge, dignity and opportunities for direct interaction (Van Wynsberghe and Comes 2020).

Critical work similarly warns that so-called "humanitarian AI" may reduce care and communication to its barest instrumental forms, and thereby ***de-humanizing*** the relationship between affected communities and aid agencies. This disconnection is compounded by the extraction of value from collected data or experimentation for the purposes of audit and metrics, and risks reproducing colonial legacies of humanitarian aid whilst obscuring the power dynamics at play (Madianou 2021).

| Key concept | **Crisis imaginary:** a shared set of assumptions about what a crisis/disaster is, how it develops, which events are seen as 'threats', and which interventions are seen as legitimate or necessary. HST may contribute to an imaginary in which disaster is a data-rich and rapidly evolving environment that demands continuous sensing, anticipation and algorithmic optimisation. Such an imaginary can be valuable when earlier detection and better coordination helps to save lives, but it can also narrow the range of questions asked about crisis, e.g. shifting the attention from the political distribution of vulnerability toward the technical management of its symptoms. |
|---|---|

## Changing Boundaries of Intervention

Technologies rarely remain within the contexts for which they were initially developed. The concept of **dual-use** is often understood narrowly as the possibility that a technology developed for beneficial purposes (e.g.

life-saving) can subsequently be used for harmful purposes (e.g. targeted killing). In practice, the migration between domains is more complex. Algorithms, datasets, expertise, procurement models, standards, organisational routines and even concepts of 'responsible action' can move between humanitarian, civilian, policing, border and military settings in which the distinction between 'beneficial' and 'harmful' use becomes blurry (Boucher 2015; Novitzky et al. 2018).

This means that dual-use should not be treated only as a question of *whether* a particular technology (e.g. drone swarm) could be weaponised. Dual-use also concerns, more broadly, the **movement of capabilities and categories across use domains**. For instance, a system designed to identify vulnerable people may later be useful for identifying 'suspicious' behaviour or to track down easy or weak targets ("quick wins"); a communications architecture developed for emergency coordination may become useful for crowd management; and data collected for disaster mapping may acquire value for national security purposes. A concept such as "vulnerability assessment" may serve different political functions when applied to populations affected by disaster, migrating people at borders or enemy targets in a military ISR missions.

HST may contribute to the **blurring of the civil-military divide** in two directions: (i) on the one hand, coercive technologies (e.g. weaponized drones) tested in exceptional contexts can return to domestic policing under narratives of urban emergency—thereby letting militarized assumptions become 'ordinary' public-safety infrastructures (O. Davis 2019); (ii) on the other hand, a system developed for flood mapping may later support border surveillance, while a rescue-oriented classification may become a security threat indicator—thereby letting civil innovation lower the political and economic costs of magnifying military capabilities (Pallister-Wilkins 2015). The risk at stake here is that surveillance, medical monitoring, and militarized assumptions gradually become ordinary features of public protection. Ethical assessment must therefore consider foreseeable trajectories and changing interpretations, not only declared present purposes, and must thereby be informed by both **interdisciplinary and public involvement** (Taebi et al. 2014).

## Changing Values

In order to enable ethics to play a more productive role in technology development, rather than merely voicing critiques after design and harm, scholars have developed approaches such as **Value-Sensitive Design (VSD)** and **Responsible Innovation (RI)**. These frameworks rest on the assumption that no technology is neutral, and that every technical choice embeds specific social, moral and/or political values. In order to prevent arbitrary or harmful embedding of values, VSD and RI offer methodologies for systematically accounting for (in)direct stakeholders by iteratively integrating human values (e.g. privacy, fairness, dignity) into technical specifications from conceptual stages through deployment (J. Davis and Nathan 2015).

At the same time, scholars have warned that such approaches wrongly assume that values are "pre-existing" and "stable" requirements that can simply be identified and embedded into a technical system. The literature on **technomoral change** challenges this assumption and claim that technology and values *co-evolve*: moral frameworks guide technological development, but technological changes can in turn alter moral practices, expectations and value priorities (Swierstra 2013; Kudina and Verbeek 2019). This value change can be disruptive in different ways: it may produce immediate uncertainty about how to interpret a new situation, it may trigger collective moral inquiry and controversy, or it may quietly stabilize into new routines without much public debate (Nickel et al. 2022). The key point is that technologies do not simply embed pre-existing values easily; they generate interpretive gaps, force reinterpretations of what privacy, security or life 'is', and create new moral questions for different groups at different times. For HST, this is particularly significant because the technology may alter not only what responders and institutions *can* do but what they increasingly feel they *ought* to do. For instance, if autonomous systems can operate continuously without getting 'tired', institutions may come to expect *continuous preparedness* and may over-value anticipatory modes of protection compared to before.

## How We Talk About HST

Projects that develop HST face ethical and societal questions not only in the development, implementation and use of the technology, but also in their **communication** about it. Talk about a technology is never merely descriptive; it helps determine what the technology is understood to be, which problems it appears capable of solving, and in what kind of 'world' it imagines it to perform. This communication includes not only press releases and public-facing outputs, but also research output, protocols, interface labels, scenario descriptions, and policy briefs. These forms of language help define who appears as a beneficiary, victim, threat, operator, or legitimate decision-maker. In this sense, communication and storytelling contributes to the ethical and political meanings of HST itself (Reijers and Coeckelbergh 2020; Coeckelbergh 2021).

A specific risk is that particular ways of describing and legitimizing HST will eliminate proper ethical inquiry. This can happen through so-called **adiaphorization**: the process by which an action is removed from ordinary moral evaluation by being presented as a neutral, technical, or necessary procedure. For instance, in the political and public discourse around drones, the targeted killing of suspected terrorists by drones is repeatedly justified in organic-medical metaphors: terrorism as ‘cancer’, drone strikes as ‘surgical’ interventions that excise diseased cells for the health of the body politic (Schwarz 2019). These narratives recode violence as a kind of therapeutic procedure, a technical operation to restore security rather than a political decision to kill. Such language directs attention towards calibration, efficiency and technical performance, while occluding questions about institutional power, competing values and human rights. The same dynamic need not involve physical violence: mechanisms of surveillance, classification, exclusion and population management can also be morally neutralised when they are presented primarily as the outputs of objective technical systems.

**Humanitarian language** can produce similar effects, even when the underlying aims are well-intended and benevolent. Describing HST as “life-saving”, “life-protecting”, or as a means of reaching communities at the “last mile” can accurately capture important benefits, particularly when human responders face dangerous or inaccessible environments. Yet such language also creates a **moral halo** around the technology. It foregrounds speed, coverage and logistical efficiency while potentially obscuring trade-offs involving privacy, dignity, contextual knowledge, de-skilling, data extraction and unequal power relations (Wang et al. 2021; Van Wynsberghe and Comes 2020; Madianou 2021). In discourse on domestic disaster-drone use, the drone is often presented as a heroic, life-saving actor, while the political causes of vulnerability and the responsibilities of public institutions become less visible (Schnepf 2020). The point is not that these aims/claims are insincere, but that they can make criticism of HST appear to be criticism of assistance itself. Questioning *how* assistance is delivered can then be misrepresented as opposing the provision of assistance.

Such narratives do not merely describe autonomous technologies; they help establish which forms of autonomy appear acceptable and which remain politically unacceptable, thereby shaping the range of policy options considered realistic (Ferl 2024). This raises an important distinction: while **public acceptance** concerns whether/to what extent people appear willing to tolerate or support the use of a technology, **public legitimacy** asks a more demanding question: can the technology and the arrangements governing it be justified to those affected by it, *including* those who object to or contest it? Strategies for civilian drone development have sometimes emphasized S&R and other benevolent applications in order to reduce public concerns about military associations and controversial uses (Boucher 2015). Such strategies may increase public acceptance *without* resolving whether people have been adequately informed about dual-use pathways, surveillance capacities or possibilities for future repurposing. A technology can therefore become ‘acceptable’ through selective framing while remaining democratically under-justified.

## Design Implications for the Imaginary

| Implication | What this means in practice | Who should act |
|---|---|---|
| **1. Design HST around explicit assumptions about crisis/disaster** | All technical and operational development should make explicit what HST is expected to detect, classify and prioritise, and what it is not designed to capture. System requirements and roadmaps should avoid treating what can be measured as equivalent to what is most important to respond to. | Developers; procurers; scenario designers |
| **2. Design for responsible capability migration** | HST should be developed with attention to how its data, algorithms, infrastructures, expertise and operational procedures could be reused beyond the immediate disaster-response context. Data architectures, access rights, interfaces and system specifications should support appropriate purpose limitation and traceability where feasible. | Developers; funders; procurement and export-control authorities |
| **3. Maintain discursive reflexivity in communications** | Scholarly, public and policy communications about HST should avoid euphemistic framing and precision metaphors. Use language that keeps moral questions visible (e.g. 'human dignity', 'shared decision', 'uncertainty') rather than only 'autonomous operation', 'precision' or 'algorithmic optimisation'. | Researchers; communication officers; policymakers |
| **4. Maintain ethics as ongoing dialogue, not compliance** | Ethical work in technology projects should be a sustained critical conversation with engineers and end users, not a one-time sign-off. Build in moments of collective ethical reflection at each major development milestone, and integrate ongoing ethical reflection in training. | Research consortia; funders; training institutes |

SECTION 4

# 4. Ethics at the Joints: Two Problems

Sections 2 and 3 discussed the three levels one at a time. This section shows why that is not enough. It follows two problems through all three levels and shows how each changes shape as it moves through the translations. In both cases, a fix at one level alone fails, because the problem reappears at the next joint.

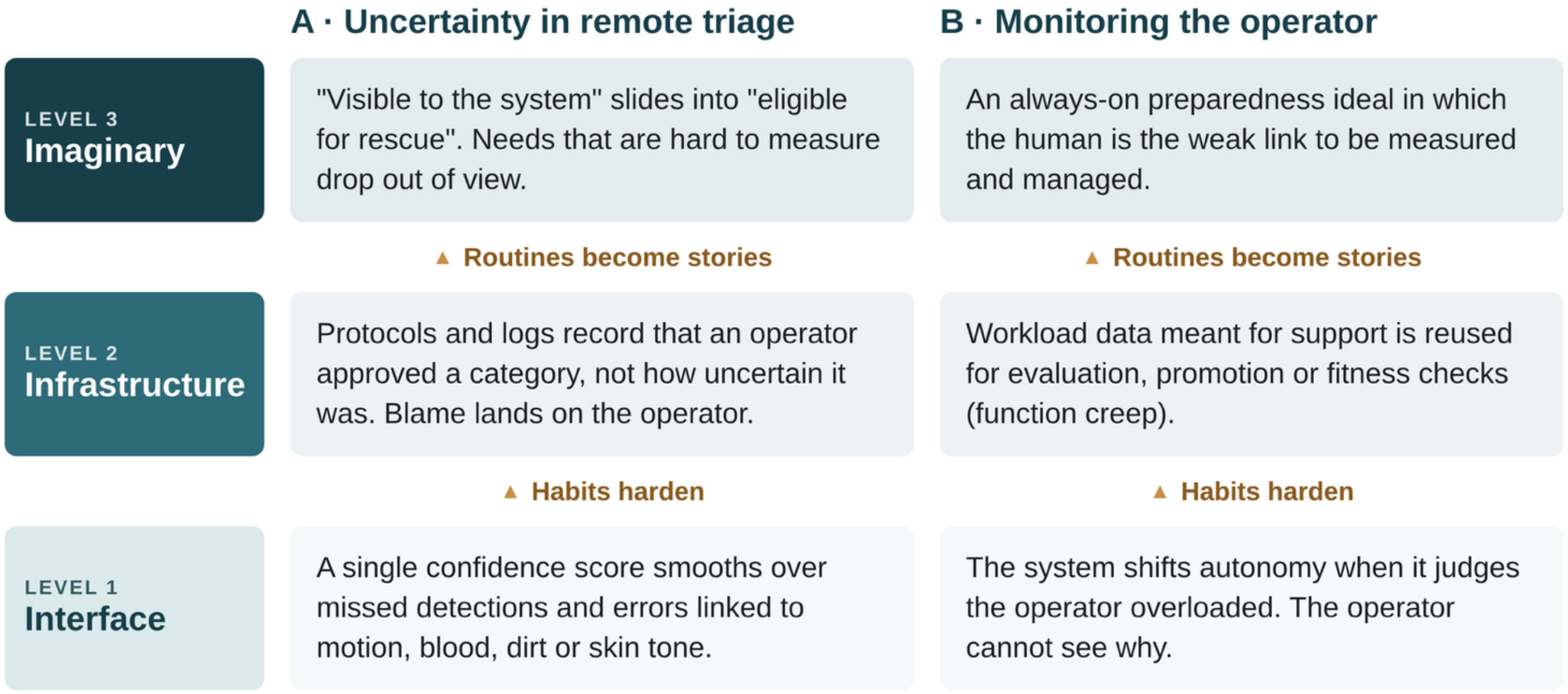

**Figure 3.** Two problems followed across the three levels. The upward arrows mark the translations through which each problem moves; the reverse translations (principles become procedures, procedures become prompts) push it back down.

## 4.1 Uncertainty in Remote Triage

**At the Interface**, a remote triage system presents each casualty with a category and a confidence score. Current prototypes struggle with motion, occlusion, blood, dirt and skin tone, and some fail in a substantial share of recordings (Section 3.2). A single score smooths these differences away. The operator sees a decision to approve, not the uncertainty behind it.

**Habits harden.** Organisations build protocols around what the interface offers. Logs record that an operator approved a category, not how uncertain the category was or what the operator could have known. When something goes wrong, the record points to the operator: the moral crumple zone.

**Routines become stories.** As remote triage becomes routine, “being visible to the system” can quietly become “being eligible for rescue”. People who are hard to detect, such as those under debris, indoors or outside sensor range, drop out of the operational picture and, eventually, out of the public account of the disaster.

**Why a single-level fix fails.** Better uncertainty displays alone do not help if protocols still only log approvals. Better logs do not help if the public story equates detection with need. Each level needs its own measure: interfaces that show uncertainty and occluded zones; protocols that record the reasons available at the time; and public accounts that report what the system could not see.

## 4.2 Monitoring the Operator

**At the Interface**, adaptive autonomy estimates the operator's workload or fatigue from physiological signals and shifts tasks to the swarm when it judges the operator overloaded. The operator often cannot see why the shift happens or how their own body is being read (Section 3.1). Models trained on neurotypical operators may misread others.

**Habits harden.** Data collected to support operators becomes available to the organisation. It can drift into evaluation, promotion or fitness decisions: function creep. Because consent is weak in hierarchical organisations, the line between support and surveillance is drawn by the infrastructure, not by the operator.

**Routines become stories.** At the level of the Imaginary, the monitored operator fits an always-on ideal of preparedness, in which the human is the weak link to be measured and managed. This in turn makes further automation look like the natural remedy.

**Why a single-level fix fails.** Making autonomy shifts visible at the interface is necessary but not enough. Without purpose limitation in the infrastructure, the same data can still be repurposed. Without questioning the preparedness ideal, each new safeguard can be read as one more reason to monitor. Responsible design here pairs transparent adaptive autonomy with strict data governance and an explicit account of what human judgment is for.

Both examples show the same pattern. Problems that look technical at the Interface become organisational in the Infrastructure and political in the Imaginary. The reverse translations work the other way: principles become procedures, and procedures become prompts on the operator's screen. This is why the design implications in Section 3 should be read together, not as three separate checklists.

SECTION 5

# 5. Glossary of Key Ethical and Legal Concepts

This glossary defines the key concepts used in this report. Entries are organised by level and include a brief example.

## The Interface (Micro-Level Concepts)

**Technological mediation**: The way technologies shape human perception and moral judgment by amplifying some features of a situation and attenuating others, rather than simply presenting reality neutrally. *Example:* whether a swarm interface highlights thermal signatures or de-emphasises uncertainty is not a neutral display choice; it shapes what operators can notice and care about.

**Evidential ecology**: The mediated field of information an HST interface produces: what is visible, what counts as evidence, and what courses of action are implied. Design choices about visualisation and classification are simultaneously ethical choices. *Example:* the layered map, icons, alerts and confidence scores on a swarm dashboard.

**Interpretive labour**: The cognitive and moral work operators perform to make sense of complex, algorithmically produced outputs, going beyond "decision-making" to include sense-making, uncertainty management, and judgment under time pressure. *Example:* studies of operator workload should measure the quality of interpretation, not only task performance.

**Response-ability**: The capacity of operators to remain perceptually and morally attuned to their situation, distinct from "control" (continuous command) or "responsibility" (post-hoc attribution). Designing for response-ability means sustaining judgment, not replacing it. *Example:* the capacity that interface design and training should protect.

**Moral deskilling**: The erosion of the skills and confidence needed for responsible moral agency, caused by systems that consistently shift decision-making away from humans in morally sensitive cases. *Example:* adaptive autonomy that routinely takes over precisely where judgment matters most.

**Moral injury**: Psychosocial harm operators may experience through guilt, shame, moral failure, or betrayal, independent of whether they exercised good judgment or violated any rule. Distance and autonomy (e.g. BVLOS operation) cannot be assumed to prevent this—mediated intimacy with victims can compound rather than reduce moral weight. *Example:* operators who watch casualties up close on a screen without being able to help.

## The Infrastructure (Meso-Level Concepts)

**Responsibility gap**: A situation where harmful outcomes occur but cannot be attributed to any single identifiable agent, because agency is distributed across designers, data, training, deployment and use. Santoni de Sio & Mecacci distinguish four types: *culpability* gaps (no one can be blamed), *moral* and *public accountability* gaps (no one can explain the "why"), and *active responsibility* gaps (no one is positioned to prevent harm). *Example:* a missed casualty for which no designer, operator or commander can be held to account.

**Moral crumple zone**: The phenomenon where the human nearest the system (typically the operator) absorbs blame for failures actually produced by the wider sociotechnical system, especially when "human-in-the-loop" requirements amount to rubber-stamping. *Example:* an operator blamed for approving a triage category they had no means to check.

**Meaningful Human Control (MHC) / Appropriate Human Judgment (AHJ)**: Not simply "human in the loop," but a property of sociotechnical systems designed to remain responsive to human moral reasons across design, training, deployment and review. Requires two conditions: **tracking** (system behaviour responds to relevant facts and human reasons) and **tracing** (a responsible human or institution understands the system's capabilities and effects). AHJ emphasises that the *right kinds* of judgment—not necessarily all decisions—must stay with humans. *Example:* human control over mission purpose, operating boundaries and morally significant classifications, rather than over every drone movement.

**CONTESTABILITY**: The requirement that MHC include real mechanisms for operators, affected communities, and oversight bodies to challenge system outputs, mission purposes, or institutional decisions — not just receive an explanation. *Example:* an operator challenging an automated priority, or an affected community challenging how triage categories are assigned.

**DISTRIBUTED RESPONSIBILITY**: A model in which responsibility is shared, traceable, and actionable across multiple actors and lifecycle phases, rather than concentrated in one operator. Often operationalised as **role-sensitive accountability matrices** linking swarm actions back to interface engineers, operators, commanders, and organisations. *Example:* a role-based accountability matrix linking swarm actions to interface engineers, operators, commanders and organisations.

**TRANSPARENCY, TRACEABILITY, EXPLAINABILITY (TTE)**: Three related but distinct properties: *transparency* (workable understanding of how the system operates), *traceability* (post-hoc capacity to audit behaviour), and *explainability* (contextually appropriate, just-in-time reasons for actions). Explanation without a reviewer or remedy risks a **transparency fallacy** — appearing accountable without giving anyone genuine capacity to verify or contest a decision. *Example:* logs that record what an operator could see when approving a decision, not only the decision itself.

## The Imaginary (Macro-Level Concepts)

**CRISIS IMAGINARY**: A shared set of assumptions about what a crisis is, how it unfolds, and what interventions are legitimate. HST may foster an imaginary in which disaster is primarily a data problem requiring continuous sensing and optimisation — which can help save lives, but can also narrow attention away from the political distribution of vulnerability. *Example:* the assumption that a disaster is primarily a data problem that more sensing will solve.

**OPERATIONAL VS. POLITICAL VISIBILITY**: Whether a system can *locate* something (operational) is distinct from whose situation becomes legible as a public problem deserving resources (political). Faster detection does not automatically produce more just disaster governance. *Example:* a flood map that locates damage accurately but leaves unequal protection out of the frame.

**DUAL-USE**: Beyond the narrow sense of "beneficial tech later weaponised," dual-use concerns the migration of *capabilities and categories* across domains: a vulnerability-detection system may be repurposed to identify "suspicious" behaviour; a rescue coordination architecture may support crowd control. This includes the blurring of civil-military boundaries in both directions. *Example:* a system for detecting vulnerable people that is later used to flag “suspicious” behaviour.

**VALUE-SENSITIVE DESIGN (VSD) / RESPONSIBLE INNOVATION (RI)**: Methodologies for systematically integrating human values (privacy, fairness, dignity) into technical design from the outset, on the assumption that no technology is value-neutral. *Example:* involving responders and affected communities in specifying what a swarm should detect and display.

**ADIAPHORIZATION**: The process by which an action is removed from ordinary moral evaluation by being framed as neutral, technical, or necessary — e.g. describing lethal operations in clinical/medical metaphors ("surgical," "clean"). Humanitarian language can produce a related effect (a "moral halo") that obscures trade-offs even when well-intentioned. *Example:* calling a swarm deployment “precise” or “surgical” in public communication.

**PUBLIC ACCEPTANCE VS. PUBLIC LEGITIMACY**: Acceptance asks whether people are willing to tolerate a technology; legitimacy asks the more demanding question of whether it can be justified to those affected by it, *including* those who object. A technology can become acceptable through selective framing without being democratically justified. *Example:* framing civil drones mainly as life-saving may raise acceptance without justifying dual-use pathways to the public.

SECTION 6

# 6. Conclusion: Lessons and Open Questions

This report has mapped the ethical and societal territory that human-swarm teaming opens up in crisis governance. Three lessons stand out:

1) **The key concerns are relational.** Problems and solutions at the Interface, in the infrastructure and in the Imaginary translate into one another. Addressing them one level at a time leaves the joints unguarded, and the joints are where responsibility gets lost.
2) **Meaningful human control is made by conditions, not by a human in the loop.** Whether control is meaningful depends on what operators can see, how uncertainty is shown, whether they can contest and override, and how organisations appraise their decisions.
3) **What a swarm makes visible is an ethical and political question.** Sensing, fusion and display decide who appears as a person in need, a responder or a threat. These decisions should be made explicitly and remain open to challenge.

## 6.1 Open Questions for Research

The analysis leaves several questions open. They set the agenda for further work, including the next stages of the research behind this report on meaningful human control, responsibility allocation, public expectations and regulation.

- What makes human control over swarms meaningful rather than procedural, and how can the difference be assessed?
- How should confidence scores and other expressions of machine uncertainty be presented so that they support, rather than replace, human judgment?
- What do contestability mechanisms assume about the operator's capacity to recognise when a system's picture or priorities have become inadequate?
- Can a fused machine picture serve as a shared operational picture for several agencies without narrowing plural judgment?
- How do civil and military uses of the same swarm technologies shape each other, and how can purpose limitation be secured?

## 6.2 Questions for Developers, Procurers and Emergency Organisations

The following questions translate the analysis into practical entry points. They were first developed as a questionnaire for the engineering and organisational work packages of STEADFAST and have been generalised here. They are not a compliance checklist. “Not decided yet” is often the most useful answer, because it marks a choice that can still be made well.

### The Interface

- How are operator states such as workload, fatigue or stress defined and measured? Does any model infer emotional states?

- What baseline of “normal” functioning does a cognitive-state model assume? Does it account for neurodivergent operators?
- Are uncertainty and confidence shown to the operator, for example for a triage category, a route or a detection? Are conflicts between sensors shown or merged?
- When a missing value is inferred, is it marked as inferred? Can the operator tell a measured casualty location from a predicted one?
- Which labels does the interface use for people (casualty, responder, helper, suspicious, adversary)? Who defines them?

## The Infrastructure

- Which levels of autonomy exist, what triggers a shift between them, and who sets the thresholds? Is there a limit to what the system may take over?
- Can operators at all times see, contest, override or stop an autonomy shift or task assignment? What does the swarm keep doing when the link to the operator is lost?
- What does the system log (outputs, actions, overrides, reasons)? Is the log meant for debugging, training, incident review or evidence?
- What data about operators is stored, for how long and for what purpose? Could it be used for evaluation or fitness checks?
- Who validates remote triage against clinical judgment, and who carries clinical responsibility for a drone-based triage category?
- Which psychosocial support is available for operators exposed to casualty footage?

## The Imaginary

- Is the same hardware and software meant to serve both civil and military uses? If so, which components differ?
- Is the same data fusion meant to locate both “victims” and “enemy forces”? How are these categories kept apart?
- What is the system not designed to detect, and who decided this?
- How are “acceptance” and “trust” defined and measured? Is there also a framework for public legitimacy, including for those who object?
- What kinds of adversary or threat justify features such as counter-jamming, and do they apply in civil settings?